\documentclass{elsart}

\usepackage{epsfig}

\usepackage{amssymb}

\def \pt {p_T}
\def \roots {\sqrt{s_{_{NN}}}}
\def \dphi {\Delta\phi}
\def \deta {\Delta\eta}

\begin{document}

\begin{frontmatter}



\title{Jet-Like Correlations between Forward and Mid-Rapidity}


\author{Fuqiang Wang (for the STAR Collaboration)}

\address{525 Northwestern Ave, Department of Physics, Purdue University, IN 47907, USA}

\begin{abstract}
% Text of abstract
Mid-rapidity azimuthal correlations probe di-jets originating mainly from gluon-gluon hard-scattering. Measurements of such correlations have revealed significant (gluon-)jet modification in central Au+Au collisions. Azimuthal correlations at forward rapidity with a mid-rapidity high-$\pt$ hadron, on the other hand, are sensitive primarily to quark-gluon hard-scattering and may probe quark-jet modification in nuclear medium. They may be also sensitive to the Color Glass Condensate by separating d-side and Au-side in d+Au collisions. We present the first results on correlations of charged hadrons at forward rapidity in the STAR FTPCs ($2.7<|\eta|<3.9$, $\pt<2$~GeV/$c$) with high-$\pt$ charged hadrons at mid-rapidity from the main TPC ($|\eta|<1$, $\pt>3$~GeV/$c$). Preliminary results from $pp$, d+Au, and Au+Au collisions at $\roots=200$~GeV are compared. Implications of the results are discussed.
\end{abstract}

\begin{keyword}
Heavy-ion \sep Jet-correlation \sep Forward rapidity \sep CGC
\PACS 25.75.-q \sep 25.75.Dw
\end{keyword}
\end{frontmatter}

% main text
\section{Introduction}
\label{intro}

Jet-like azimuthal correlations have shown significant modifications in central Au+Au collisions at RHIC due to the presence of the hot and dense medium created in these collisions~\cite{whitepapers}. This constitutes a strong evidence of jet quenching where high energy partons lose energy through interactions with the medium. The inferred medium density is sufficiently high that formation of the Quark-Gluon Plasma is plausible~\cite{whitepapers}.

Most jet-correlation measurements performed so far are at mid-rapidity where gluon-jets dominate at RHIC energies~\cite{whitepapers}. Measurements of mid-rapidity jet-correlation with forward high-$\pt$ hadrons have been performed for $pp$ and d+Au~\cite{FPD,PHENIXmuon}. 
%STAR has extensively studied jet-correlations of charged hadrons with a high transverse momentum ($\pt$) trigger particle at mid-rapidity (pseudo-rapidity $|\eta| < 1$) using the main Time Projection Chamber (TPC)~\cite{whitepapers}. 
STAR also has the capability to carry out correlation measurements at forward rapidities using the Forward Time Projection Chambers (FTPCs), which cover the range of $2.7<|\eta|<3.9$, with mid-rapidity trigger particles from the main TPC. These correlations should be dominated by quark jets, as discussed below. This paper presents the first such measurements. They potentially address two problems as follow.
%STAR has also the capability using the FTPC to measure the forward-rapidity associated particles with a high-$\pt$ trigger particle at mid-rapidity in the main TPC. 

\subsection{Jet Quenching at Forward Rapidity}

Mid-rapidity jet-correlation measurements have revealed a number of novel phenomena: (i) The away-side jet-correlation is significantly broadened in central Au+Au collisions. For some kinematic regions, the away-side correlation may even peak away from $\pi$ (opposite direction to the trigger particle in azimuth)~\cite{jetspectra,otherpapers}. (ii) Recent 3-particle azimuthal correlation results from STAR reveal structures consistent with conical emission~\cite{UleryHP06}. (iii) The near-side jet-correlation at mid-rapidity is observed to have a long range $\deta$ correlation, rather flat within the measured region of $|\deta|^{<}_{\sim}1.5$ (the so-called ridge)~\cite{PuschkeHP06}. Will these phenomena also be present at forward rapidities? How far does the ridge extend in pseudo-rapidity? These questions may be addressed by jet-correlation measurements at forward rapidities.

Gluon energy loss is predicted by Quantum Chromodynamics (QCD) to be stronger than quark's by a factor of 9/4~\cite{pQCDeloss}. To test the QCD energy loss picture, it is valuable to also have measurements where quark jets dominate, which can be obtained from measurements at forward rapidities.
%Due to the smaller energy loss rate, it is possible that jet-correlations at forward rapidities are less modified than those at mid-rapidity, given that the pathlengths of the matter traversed by the mid- and forward-rapidity partons are the same, which would be the case in the Bjorken invariance picture. On the other hand, the forward-rapidity partons may have to traverse a longer longitudinal path inside the medium than those at mid-rapidity, thus lose more energy. This pathlength effect may be more than compensating the smaller energy loss rate of quarks so that the forward jet-correlations may be more significantly modified. 
Since quarks are expected to lose less energy than gluons, one would expect that jet-correlations at forward rapidities are less modified than at mid-rapidity, where gluon fragmentation dominates. However, the pathlengths through the medium may also be different for jets at forward and mid-rapidity. Detailed calculations incorporating realistic medium denisties and expansion dynamics are likely needed to disentangle these effects.

%Recently Hwa and Yang~\cite{hwa} argued that jet-correlations may be absent from forward rapidities if particles are produced by recombination. This can be tested by our jet-correlation measurements at forward rapidities.

\subsection{The Color Glass Condensate}

The mid-rapidity di-jet correlations measured at RHIC stem primarily from hard scatterings of partons of small Bjorken $x$~\cite{whitepapers}. They are dominated by gluons because at small $x$ gluons far outnumber quarks~\cite{PDF}. On the other hand, a di-jet with one at mid-rapidity and the other at forward rapidity comes from parton scattering of quite different kinematics. While the mid-rapidity jet is still dominated by small-$x$ gluons, the forward jet has to come from a large-$x$ parton in order to be at forward rapidity with balancing $\pt$. At large $x$, valence quarks dominate~\cite{PDF}. Thus, a mid- and forward-rapidity di-jet primarily comes from hard-scattering between a small-$x$ gluon and a large-$x$ quark. For mid-rapidity trigger particle $\pt>3$~GeV/$c$ reported here, the initial gluon energy is of the order of 5~GeV, or $x_g\sim 0.05$; that for the quark is $x_q\sim x_g\sinh(\eta)\sim 0.7$.

The asymmetric parton scattering kinematics, coupled with the asymmetric d+Au collisions, provide an opportunity to probe the Color Glass Condensate (CGC)~\cite{CGC}. 
%Although the number of trigger particles (thus gluon-jets) is divided out in these measurements, the differentiation between positive and negative rapidities in d+Au allows one to assess information about the CGC. 
%If the relative number of small-$x$ gluons to large-$x$ quarks are the same in Au-nucleus and in deuteron, then there will be the same numbers of g(d)+q(Au) and g(Au)+q(d) scatterings. Normalized by the total number of trigger particles (gluon-jets), the jet-correlation signals at d-side and Au-side will be the same and equal to half of the jet-correlation from a single quark ``fragmentation". If, on the other hand, the relative number of gluons to quarks is reduced in Au-nucleus with respect to deuteron, say by a factor of $\alpha$, then the number of g(Au)+q(d) scatterings is a factor of $\alpha$ of that of g(d)+q(Au) scatterings. Normalized by the total number of gluon-jets, the probability to have a d-side quark-jet is $\alpha/(1+\alpha)$ and a Au-side quark-jet, $1/(1+\alpha)$. Thus, the jet-correlation strength at d-side is $\alpha/(1+\alpha)$ of that from a single quark fragmentation and that at Au-side is $1/(1+\alpha)$, hence the d-side signal strength will be a factor of $\alpha$ of that at the Au-side. Such a reduction at the d-side would be a signature of CGC.
If the relative number of small-$x$ gluons to large-$x$ quarks in Au-nucleus equals to that in deuteron, there will be equal numbers of g(d)+q(Au) and g(Au)+q(d) scatterings. Normalized by the total number of trigger particles (gluon-jets), the jet-correlation strength at d-side and Au-side will be the same and equal to half of that from a single quark ``fragmentation". If, on the other hand, the number of gluons is reduced relative to that of quarks in Au-nucleus, then there will be fewer g(Au)+q(d) than g(d)+q(Au) scatterings. 
%Normalized by the total number of gluon-jets, the probability to have a d-side quark-jet is reduced relative to a Au-side quark-jet. 
As a result, the d-side correlation strength will be reduced relative to that at the Au-side. Such a reduction would be a signature of CGC.

\section{Analysis and Systematic Uncertainties}
\label{analysis}

The two main detectors used in this analysis are the STAR's main TPC~\cite{TPC} and the FTPCs~\cite{FTPC}. The analysis selects high-$\pt$ trigger particles, $3<\pt^{trig}<10$~GeV/$c$, from the main TPC within $|\eta|<1$, and correlates them with charged particles measured in the FTPCs within $2.7<|\eta^{assoc}|<3.9$. The $\pt$ cut for the associated particles is $0.2<\pt^{assoc}<2$~GeV/$c$, where the upper cut is dictated by the relatively poor momentum resolution of the FTPCs. This also prevents us from selecting high-$\pt$ trigger particles from the FTPCs.

The combinatorial background is obtained from event-mixing. The elliptic flow modulation for Au+Au collisions, $2v_2^{trig}(\pt^{trig})v_2^{assoc}(\pt^{assoc})\cos(2\dphi)$, is added in pairwise during event-mixing. The trigger particle elliptic flow $v_2^{trig}$ is taken to be the average from the modified reaction plane and the 4-particle cumulant methods, and the range of the two is used as an estimate of the systematic uncertainty, as in~\cite{jetspectra}. The associated particle $v_2^{assoc}$ is measured for 20-70\% Au+Au collisions in the FTPCs by several methods which yield similar results~\cite{v2MRP}. In this analysis the two-particle cumulant $v_2\{2\}$ is used and parameterized as $v_2(\eta,\pt) = 0.0716\,(1-\exp[-(\pt/0.50)^{1.81}]) \times 2.29\,\exp[-(\eta/2.51)^2/2]$. The centrality dependence of the FTPC $v_2$ is parameterized using preliminary STAR results. 
%Although the difference in the FTPC $v_2$ results from different methods is not as large as that in the main TPC, the relative systematic uncertainty of the main TPC $v_2$ is used as a conservative estimate for the systematic uncertainty on the FTPC $v_2$ results.
The systematic uncertainty on the FTPC $v_2$ is taken, conservatively, to be as same as that on the main TPC $v_2$.

The results reported here have been corrected for tracking efficiency and acceptance of associated particles, and normalized per trigger particle. Besides $v_2$, the other major source of systematic uncertainty comes from background normalization which is described in section~\ref{results}. An additional overall systematic error of 10\% is estimated due to efficiency and acceptance corrections.

One should note that the FTPCs cover a limited range in polar angle, $8^\circ$ ($\eta=2.7$) to $2^\circ$ ($\eta=3.9$) from the beam axis. In addition, the pseudo-rapidity of the away-side jet is not known~\cite{jetspectra}. Thus, the associated particles measured in the FTPCs are only a small fraction of the away-side jet. The correlation strength is subject to this effect, however the correlation shape is not. We also note, because the FTPCs are very close to the beam axis, that some of the away-side jet fragments can sway into the near side. However, a simple toy {\em Monte Carlo} indicates that those particles do not give a near-side peak in azimuth around the trigger particle; they are more or less evenly distributed.

\section{Results and Discussions}
\label{results}

We first present and compare the forward-rapidity jet-correlation results from $pp$ and d+Au collisions. The d+Au results potentially probe the CGC. We then compare these results to Au+Au collisions. The question to address by this comparison is whether or not the energy loss and the jet quenching picture differ between mid- and forward rapidities.

\subsection{Results from $pp$ and d+Au}

\begin{figure}[htbp]
\centering
\includegraphics[width=0.75\textwidth]{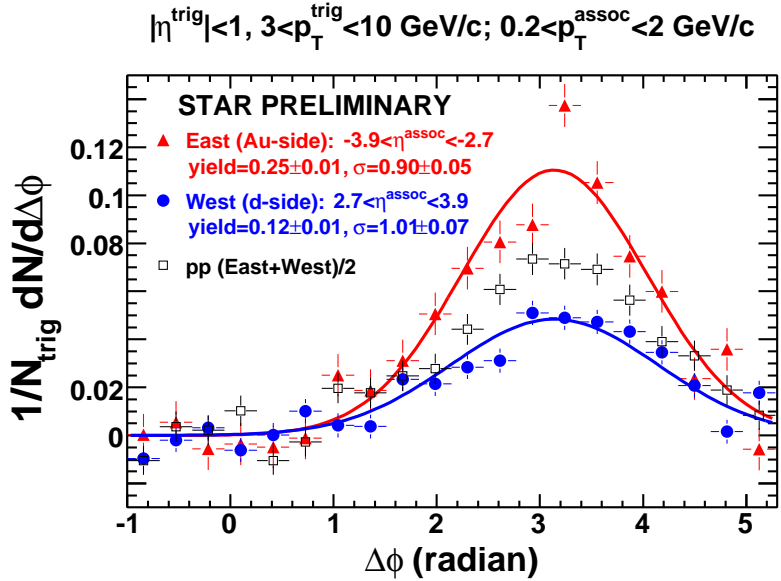}
\caption{(color online) Azimuthal correlations for the d-side (blue circles) and the Au-side (red triangles) separately in d+Au collisions, and that averaged over positive and negative pseudo-rapidities in $pp$ collisions (open squares). The subtracted backgrounds were obtained from event mixing and normalized to the signal in the entire near-side region of $|\dphi|<1$. The curves are Gaussian fits to the d+Au data.}
\label{fig1}
\end{figure}

Figure~\ref{fig1} shows the azimuthal correlations in minimum $pp$ and d+Au collisions. The positive (d-side) and negative (Au-side) pseudo-rapidity results are presented separately for d+Au collisions; the two $\eta$ regions are averaged for $pp$. The combinatorial background are normalized to the signal in the near-side region of $|\dphi|<1$ where no correlation signals are expected. (The near-side correlation is contained in the vicinity of the trigger particle which is restricted at mid-rapidity in the main TPC.) While the correlation shapes are the same, the correlation strength at the d-side is about a factor of 2 smaller than that at the Au-side. The $pp$ data is approximately equal to the average of the d- and Au-side correlation strengths. This is expected as the sum of the correlation strengths, to first order, should be equal to that from a single quark fragmentation as discussed above. Our measured reduction in the d-side correlation strength is in qualitative agreement with the observed high-$\pt$ suppression in particle production at forward rapidity by BRAHMS~\cite{brahms}.

It is well known that the parton distribution of bound nucleons are different from the ones of free nucleons, so some reduction in the correlation strength at the d-side may be expected if the small-$x$ gluons in Au are suppressed due to shadowing. However, gluons at $x\sim 0.05$ are in the anti-shadowing region according to EKS98~\cite{EKS98} -- they are more abundant in heavy nuclei than in free nucleons by about 15\%, rather scale-independent. Moreover, the quark distribution at large $x$ is suppressed in nuclei, the so-called EMC effect~\cite{EMC}; the suppression is maximal (about 20\%) at $x\sim 0.7$. Both the gluon anti-shadowing and the EMC effect should make the relative number of small-$x$ gluons to large-$x$ quarks larger in heavy nuclei than in free nucleons, the opposite to the observed reduction of the d-side correlation strength. CGC calculcations, on the other hand, predict a suppression of the gluon density in nuclei at small-$x$~\cite{CGC}. The observed reduction of the d-side correlation strength is in qualitative agreement with such a suppression, and is also in line with the mid-rapidity correlation measurement with a forward $\pi^0$ by STAR~\cite{FPD}. However, we note that the forward quark from the deuteron may suffer energy degradation (nucleon rapidity shift) traversing the Au-nucleus thickness before striking a gluon from the Au-nucleus. This could also reduce the d-side jet-correlation strength, whose magnitude, however, needs theoretical investigation.
% however, this effect cannot account for more than a factor of 2 reduction in its fragmentation multiplicity. Therefore, the data call for other suppression mechanism(s), such as the CGC; the data indicate that the $x\sim 0.05$ gluons are suppressed by about a factor of 2 in the Au-nucleus as compared to deuteron. 
\subsection{Results from Au+Au}

\begin{figure}[htbp]
\centering
\includegraphics[width=\textwidth]{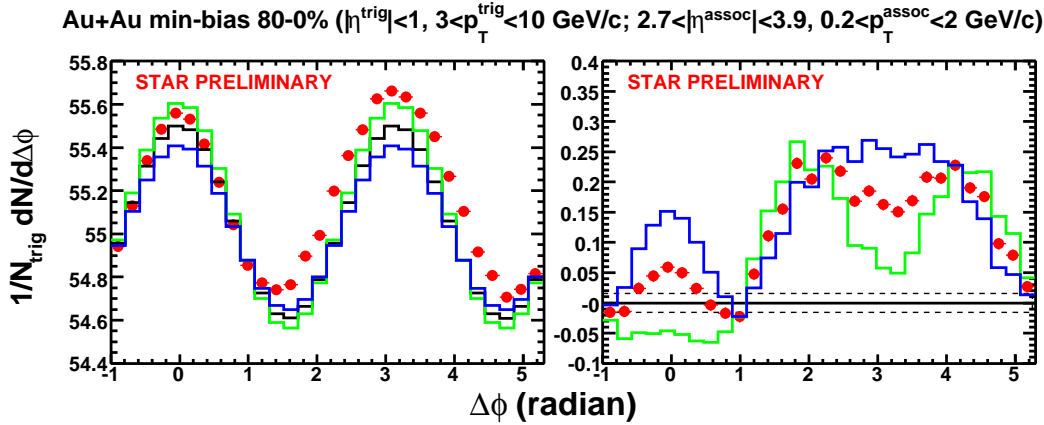}
\caption{(color online) Raw (left panel) and background-subtracted (right panel) azimuthal correlations in minimum bias Au+Au collisions. The histograms in the left panel are flow-modulated combinatorial background normalized to the signal within $0.8<|\Delta\phi|<1.2$ (black) and the corresponding systematic uncertainties (blue and green) due to elliptic flow. The histograms in the right panel are the systematic uncertainties on the correlation result from elliptic flow, and the dashed lines indicate those due to background normalization.}
\label{fig2}
\end{figure}

Figure 2 shows the azimuthal correlation results in minimum bias Au+Au collisions (about 80\% of the total geometrical cross-section). The raw correlation is shown in the left panel together with the combinatorial background. The background-subtracted result is shown in the right panel. The combinatorial background was obtained from event mixing where the elliptic flow modulation is added in pairwise. The background is normalized to the signal within the normalization range of $0.8<|\Delta\phi|<1.2$, where the minimum correlation signal is found. The systematic uncertainty on the background normalization is estimated by varying the size of the normalization range around $\dphi=1$. We do not normalize the background to the entire near-side region as for $pp$ and d+Au because the near-side correlation in Au+Au may not be zero {\em a priori}; the $\deta$ ridge observed at mid-rapidity~\cite{PuschkeHP06} could extend to forward rapidities.

As seen from the right panel of Fig.~\ref{fig2}, the near-side correlation is consistent with zero within the systematic uncertainties which are presently large. The uncertainties need to be reduced in order to address whether or not the observed mid-rapidity $\deta$ ridge extends to forward rapidities. The away-side correlation is significantly broader than that in $pp$ and d+Au collisions. Given the present systematic uncertainties, the away-side correlation is consistent with either a flat or a double-peaked distribution. However, the observed broadness of the away-side correlation is robust and suggests that the medium modification to jet-like correlations is as strong at forward rapidity as at mid-rapidity. 

\begin{figure}[htbp]
\centering
\includegraphics[width=0.345\textwidth]{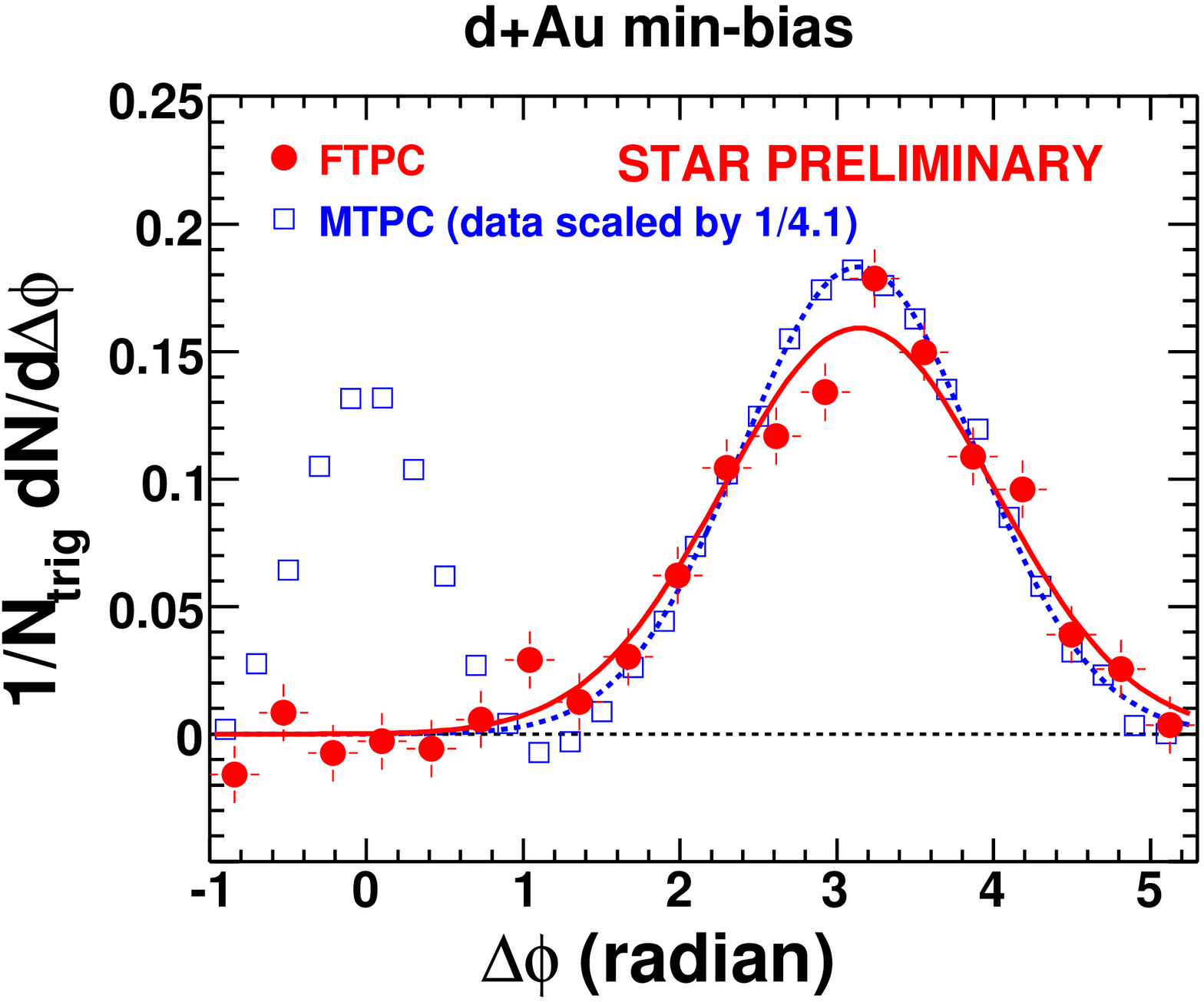}
\includegraphics[width=0.315\textwidth]{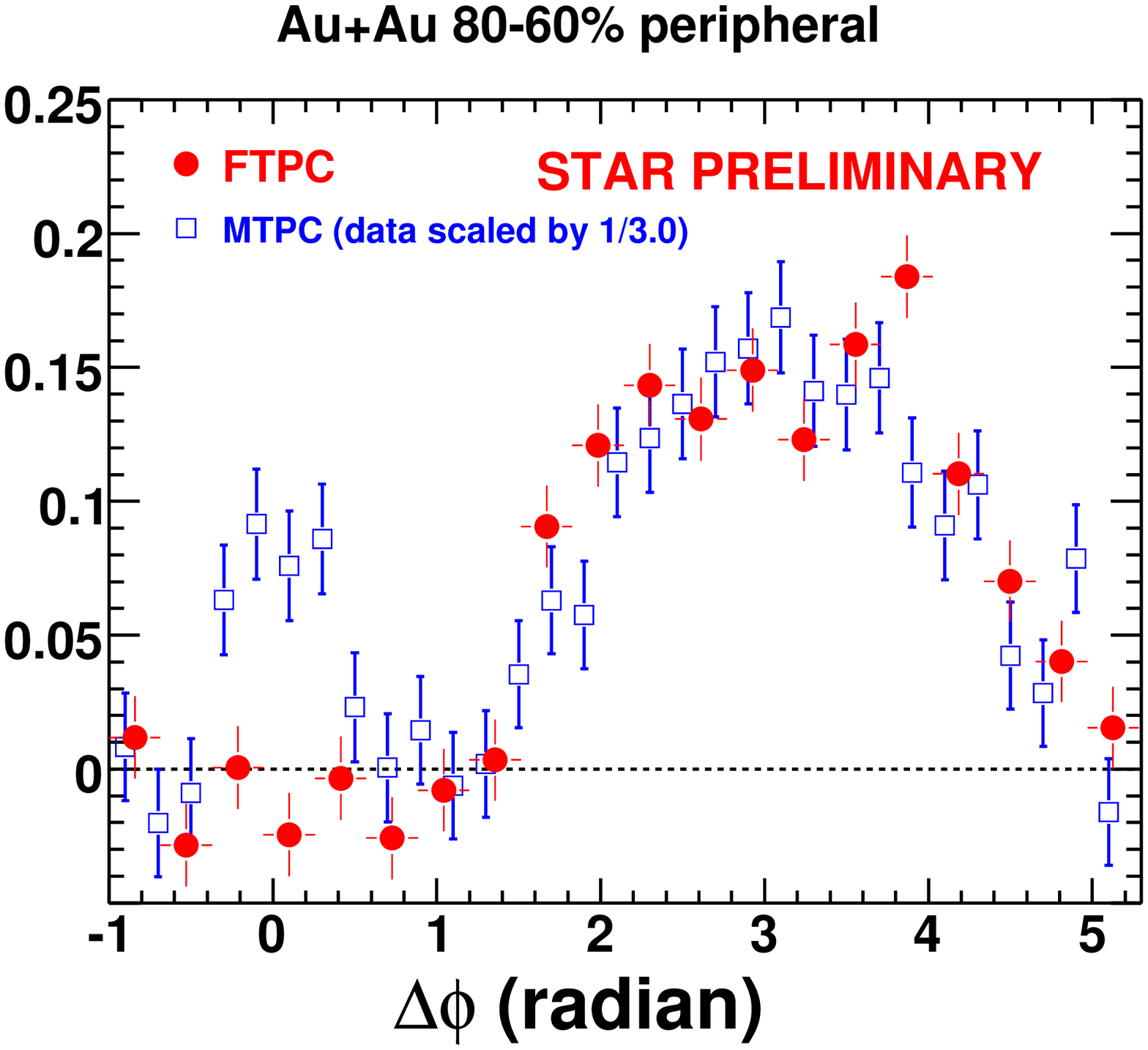}
\includegraphics[width=0.315\textwidth]{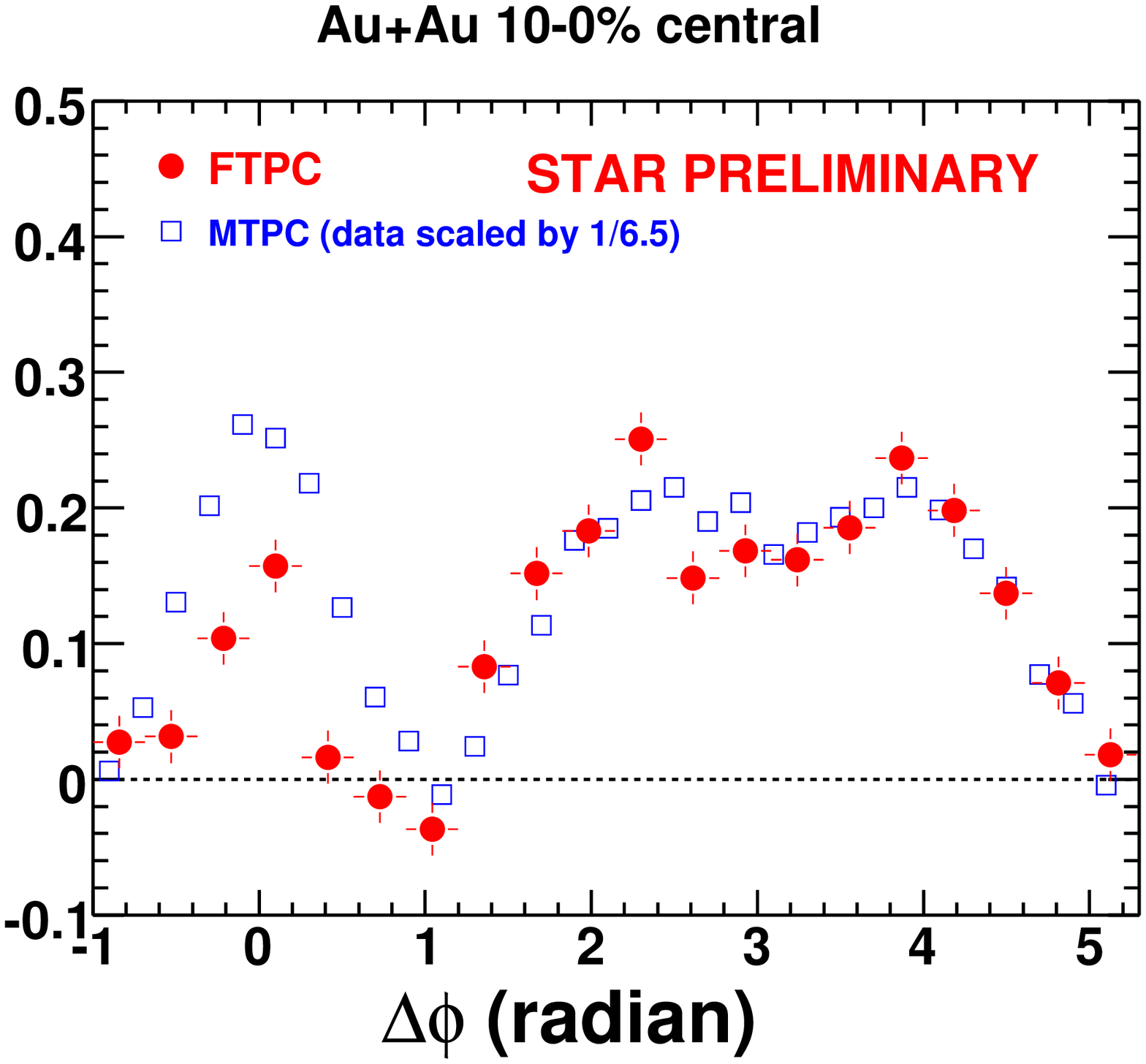}
\caption{(color online) Azimuthal correlations at forward rapidities (red) compared to mid-rapidity (blue) in d+Au (left panel), 80-60\% peripheral Au+Au (middle panel), and top 10\% central Au+Au collisions (right panel). The trigger particles are within $3<\pt^{trig}<4$~GeV/$c$ and $|\eta|<1$, and the associated particles are within $0.2<\pt^{assoc}< 2$~GeV/$c$ for forward rapidities $2.7<|\eta|<3.9$ and $0.2<\pt^{assoc}< 3$~GeV/$c$ for mid-rapidity $|\eta|<1$. The mid-rapidity results are scaled to facilitate comparison of the correlation shapes.}
\label{fig3}
\end{figure}

Figure~\ref{fig3} compares the shapes of the azimuthal correlations at mid- and forward rapidities. The near-side correlations are different because the near-side correlation centers around the trigger particle which is at mid-rapidity in the main TPC and should be naively absent from forward rapidities (except for possibly the ridge). The away-side correlations are remarkably similar between mid- and forward rapidities. One should note that the individual correlation functions have large systematic uncertainties, but these uncertainties (mainly due to flow) are strongly correlated. Therefore, while the particular shape is subject to a large uncertainty, the comparison between the shapes is rather robust. The similarity in the away-side correlation shapes suggests that the energy loss picture may be the same between mid- and forward rapidity. This could imply a number of scenarios: (1) the energy loss is so large that the partons lose almost all their energy and the different energy loss rates for gluons and quarks are insignificant; (2) the forward-rapidity partons (mostly quarks) traverse more matter (longitudinally) than the mid-rapidity partons (mostly gluons), almost perfectly cancelling the difference in their energy loss rates; and (3) the physics may just be the same between gluon- and quark-jets and between mid- and forward rapidities. Futher studies are needed in order to discriminate these different scenarios. For instance, Cu+Cu collisions and more peripheral Au+Au data may shed more light on (1) and moving to other forward rapidities may give insight into (2).

\section{Conclusions}
\label{conclusion}

We have reported azimuthal correlations of charged hadrons at forward rapidities ($2.7<|\eta|<3.9$, $\pt<2$~GeV/$c$) measured by the STAR FTPCs with high-$\pt$ charged trigger particles at mid-rapidity ($|\eta|<1$, $\pt>3$~GeV/$c$) measured by the main TPC. Preliminary results are reported for $pp$, d+Au, and Au+Au collisions at $\roots=200$~GeV. Near-side correlations are absent from $pp$ and d+Au collisions. The systematic uncertainties in Au+Au collisions do not allow a firm conclusion on the existence of near-side correlations or whether or not the observed mid-rapidity long range correlation in $\deta$ extends to the measured forward rapidities. 

Away-side jet correlation signals are observed at forward rapidities. While the correlation shapes are similar, the correlation strength at the d-side is a factor of 2 smaller than that at the Au-side in d+Au collisions, and the $pp$ result is the average of the two. The gluon anti-shadowing and the EMC effect would yield an enhanced d-side correlation.
%Thus the data suggest other mechanism(s), such as the Color Glass Condensate, be responsible for the factor of 2 suppression of small-$x$ gluons in the Au-nucleus.
The data are in qualitative agreement with small-$x$ gluon suppression in the Au-nucleus by the Color Glass Condensate. 
Quantitative calculations of the Color Glass Condensate (and possibly other physics mechanisms) are needed in order to further our understanding.

The away-side correlation in Au+Au collisions at forward rapidity broadens with centrality. The correlation shapes are similar to those at mid-rapidity. The similarity suggests a similar energy loss picture at mid- and forward rapidities. The underlying physics mechanisms require further investigations.

%\section*{Acknowledgement}
%The author would like to thank Dr. Xin-Nian Wang for the kind invitation.

% \appendix

% \section{}
% \label{}

\end{document}